\documentclass[PAPER, cernpreprint, atlasdraft=false, UKenglish, texmf, orcidlogo]{atlasdoc}
\usepackage{atlaspackage}
\usepackage{atlasbiblatex}

\usepackage{atlasphysics}
\graphicspath{{logos/}{figures/}}

\usepackage{ANA-EXOT-2024-32-PAPER-defs}

\usepackage{hhline}
\usepackage{longtable}
\usepackage{float}

\usepackage{multirow}

\newcommand\Tstrut{\rule{0pt}{2.6ex}}         %
\newcommand\Bstrut{\rule[-0.9ex]{0pt}{0pt}}   %

\AtlasTitle{Search for quantum black holes in lepton+jet final states using
proton--proton collisions at $\sqrt{s}=13.6~\textrm{TeV}$ with the ATLAS detector}

\AtlasVersion{2.0}

\AtlasAbstract{A search for quantum black holes in electron+jet or muon+jet final states with high invariant mass is performed. The analysis uses data from $\sqrt{s}=13.6~\textrm{TeV}$ $pp$ collisions recorded by the ATLAS detector between 2022 and 2024 during Run~3 of the Large Hadron Collider, corresponding to an integrated luminosity of $164~\mathrm{fb}^{-1}$. This search is strongly motivated by a dramatic increase of the production cross-section by up to an order of magnitude for the highest masses considered, thanks to the small increase of $0.6~\textrm{TeV}$ in centre-of-mass energy between Run~2 and Run~3. No significant excess above the Standard Model background is observed, and 95\% CL upper limits are set on the production cross-section times branching ratio in several benchmark models, reaching a mass scale of $9.4~\textrm{TeV}$. These represent the strongest exclusion limits to date on quantum black hole production.}

\AtlasRefCode{EXOT-2024-32}

\arXivId{2604.19495}

\HEPDataRecord{3147718}

\AtlasJournal{PLB}
\AtlasJournalRef{\PLB 881 (2026) 140849}
\AtlasDOI{10.1016/j.physletb.2026.140849}

\AtlasCoverEgroupAnalysisTeam{atlas-EXOT-2024-32-analysis-team@cern.ch}

\hypersetup{pdftitle={ATLAS document},pdfauthor={The ATLAS Collaboration}}

\begin{document}

\atlasdatapolicy{athenatext=false}

\maketitle

\section{Introduction}
\label{sec:intro}

The hierarchy problem---the immense gulf between the electroweak scale and Planck scale---may be resolved by theories with extra dimensions (EDs) that lower the fundamental $D$-dimensional scale of gravity, $M_{\textrm{D}}$, to the TeV range.
In the Arkani-Hamed, Dimopoulos, and Dvali (ADD) model~\cite{Nima, Antoniadis:1998ig}, gravity propagates through $n$ flat EDs (where $D=4+n$), while Standard Model (SM) fields are restricted to
four-dimensional space-time. Alternatively, the Randall--Sundrum (RS) model~\cite{Randall} employs one warped extra dimension to generate this scale hierarchy.
These theories predict a plethora of new phenomena, some of which are accessible at the Large Hadron Collider (LHC).
One such prediction is the production of quantum black holes (QBHs)~\cite{Gingrich:2009hj, Meade:2007sz, Calmet:2008dg, Gingrich:2009da, Gingrich:2015yda} at threshold masses $M_{\textrm{th}} \sim M_{\textrm{D}}$.
In this benchmark framework, it is postulated that the QBH parton--parton cross-section follows a power-law increase with partonic centre-of-mass energy, with the power determined by the number of extra dimensions.
For higher proton-beam energy, the suppression of the partonic cross-section arising from the parton distribution functions (PDF) at high Bjorken-$x$ (the fraction of the proton's momentum carried by the interacting parton) is reduced, resulting in a significant increase in the QBH production cross-section, $\sigma_{\textrm{QBH}}$.
Unlike semi-classical black holes, which decay thermally via Hawking radiation into high-multiplicity final states~\cite{Anchordoqui:2001cg, ATLAS:2015yln, ATLAS:2018rvc, CMS:2017boz, CMS:2018ozv, CMS:2026dqz}, these QBHs decay into two-particle final states, where lepton- and baryon-number conservation may be violated.
This search focuses on diquark production of QBHs that can decay to lepton+antiquark $(\ell+\bar{q})$ final states:
\begin{equation}
uu \rightarrow \textrm{QBH}^{4/3}\rightarrow \bar{d}\ell^{+}, \quad
ud \rightarrow \textrm{QBH}^{1/3}\rightarrow \bar{u}\ell^{+}, \quad
dd \rightarrow \textrm{QBH}^{-2/3}\rightarrow \bar{d}\ell^{-},
\label{eq:QBH_processes}
\end{equation}
where the superscript indicates the QBH electric charge and the processes are sorted by descending $\sigma_{\textrm{QBH}}$, left-to-right.
The notation $u$ and $d$ marks the up-type and down-type quarks.
The branching ratios for the processes ordered as in Eq.~(\ref{eq:QBH_processes}) are 11\%, 5.6\% and 6.7\%, to each lepton flavour~\cite{Gingrich:2009hj}.

A key feature that sets this model apart from virtually all other new-physics scenarios, and motivates this search already with a partial run dataset, is the production cross-section's strong dependence on the proton--proton centre-of-mass energy, $\sqrt{s}$.
This dependence causes the increase in $\sqrt{s}$ from Run~2 of the LHC ($13~\textrm{TeV}$) to Run~3 ($13.6~\textrm{TeV}$) to enhance $\sigma_{\textrm{QBH}}$ by approximately $100\%$ at $M_{\textrm{th}} = 6~\textrm{TeV}$ up to about $1000\%$ at $M_{\textrm{th}} = 10.5~\textrm{TeV}$~\cite{Gingrich:2009hj}.
This analysis succeeds the previous ATLAS search performed at $\sqrt{s} = 13~\textrm{TeV}$ using the full $140~\textrm{fb}^{-1}$ Run~2 dataset~\cite{ATLAS:2023vat}.
The same search was done by ATLAS at $\sqrt{s} = 8~\textrm{TeV}$ with $20.3~\textrm{fb}^{-1}$ of data~\cite{ATLAS:2013wgh}.
Complementary searches were also performed in dijet, dilepton and photon+jet final states by ATLAS~\cite{ATLAS:2015nsi,
ATLAS:2016loq, ATLAS:2019fgd, ATLAS:2013ylb, ATLAS:2015esi, ATLAS:2017dpx} and CMS~\cite{CMS:2017caz, CMS:2018hnz, CMS-EXO-19-014,CMS:2023twl}.
In comparison with the other final states, the lepton+jet ones benefit from significantly lower backgrounds, which compensates for their smaller cross-section times branching ratio.
Lepton+jet final states are of course not unique to QBH production, and other searches probing similar regions of phase space do not necessarily provide an explicit interpretation in terms of QBHs.
For example, ATLAS has published a lepton+jet search for leptoquarks with a combination of the full Run~2 dataset and a partial ($56~\textrm{fb}^{-1}$) Run~3 dataset~\cite{ATLAS:2025upm}.

The present search uses $164~\textrm{fb}^{-1}$ of $\sqrt{s} = 13.6~\textrm{TeV}$ proton--proton collision data, recorded by the ATLAS detector during the 2022--2024 period of Run~3. Even with a Run~3 dataset similar in size to the full Run~2 dataset, this analysis yields considerably higher QBH exclusion limits thanks to the significant increase in $\sigma_{\textrm{QBH}}$.
Electron+jet and muon+jet final states with an invariant mass above $3~\textrm{TeV}$ are targeted separately in this search.
Explicit contributions from QBH decays with $\tau$-leptons are not considered, since they are expected to be negligible compared to those from the other two lepton+jet final states, owing to the $\tau \rightarrow \ell \bar{\nu}_{\ell}\nu_{\tau}$ branching ratio and the softer electron and muon transverse momentum ($p_{\textrm{T}}$) spectra due to the presence of neutrinos in $\tau$ decays.
The main background processes include $V$+\;\!jets ($V=W,Z$), $VV$ and events with top quarks.
These are modelled using Monte Carlo (MC) simulations.
Events containing non-prompt electrons, or electrons from photon conversions or jets misidentified as electrons, collectively referred to as \enquote{fakes}, are estimated using data-driven techniques.
This is the dominant background in the electron channel, while $W$+\;\!jets production contributes the most background in the muon channel.

%


%

%
\section{ATLAS detector}
\label{sec:detector}

The ATLAS experiment~\cite{PERF-2007-01,GENR-2019-02} at the LHC is a multipurpose particle detector
with a forward--backward symmetric cylindrical geometry and a near \(4\pi\) coverage in
solid angle.\footnote{\AtlasCoordFootnote}
It consists of an inner tracking detector surrounded by a thin superconducting solenoid
providing a \qty{2}{\tesla} axial magnetic field, electromagnetic and hadronic calorimeters, and a muon spectrometer (MS).
The inner tracking detector covers the pseudorapidity range \(|\eta| < 2.5\).
It consists of silicon pixel, silicon microstrip, and transition radiation tracking detectors.
Lead/liquid-argon (LAr) sampling calorimeters provide electromagnetic (EM) energy measurements
with high granularity within the region \(|\eta|< 3.2\).
A steel/scintillator-tile hadronic calorimeter covers the central pseudorapidity range (\(|\eta| < 1.7\)).
The endcap and forward regions are instrumented with LAr calorimeters
for EM and hadronic energy measurements up to \(|\eta| = 4.9\).
The MS surrounds the calorimeters and is based on
three large superconducting air-core toroidal magnets with eight coils each.
The field integral of the toroids ranges between \num{2.0} and \qty{6.0}{\tesla\metre}
across most of the detector.
The MS includes a system of precision tracking chambers up to \(|\eta| = 2.7\) and fast detectors for triggering up to \(|\eta| = 2.4\).
The luminosity is measured mainly by the LUCID--2 detector which is located close to the beampipe~\cite{LUCID2}.
A two-level trigger system was used to select events~\cite{TRIG-2016-01,TRIG-2022-01}.
The first-level trigger is implemented in hardware and used a subset of the detector information
to accept events at a rate close to \qty{100}{\kHz}.
This is followed by a software-based trigger that
reduced the accepted rate of complete events to \qty{3}{\kHz} on average,
depending on the data-taking conditions.
A software suite~\cite{SOFT-2022-02} is used in data simulation, in the reconstruction
and analysis of real and simulated data, in detector operations, and in the trigger and data acquisition
systems of the experiment.


%
%
%
%
%
%
%
%
%

%

%
\section{Data and simulation}
\label{sec:data_mc}

Data and simulated event samples used in this study must satisfy a requirement that all detector systems operate normally~\cite{DAPR-2018-01}. %
The data were collected with single-electron or single-muon triggers~\cite{TRIG-2022-01}.
The $p_{\textrm{T}}$ threshold is $26~(24)~\textrm{GeV}$ for isolated electron (muon) triggers and $60~(50)~\textrm{GeV}$ for non-isolated triggers.
As the analysis targets very high mass QBHs whose decays produce a lepton with $p_{\textrm{T}} > 150$~GeV, virtually all events were triggered by high-$p_{\textrm{T}}$ lepton triggers without isolation requirements whose $p_{\textrm{T}}$ threshold is $140~(50)~\textrm{GeV}$ for electrons (muons).
The application of data-quality requirements~\cite{DAPR-2018-01} results in a sample corresponding to an integrated luminosity of $164\pm 3~\textrm{fb}^{-1}$.

QBH MC signal samples were generated using the \textsc{QBH\,v3.02} generator~\cite{Gingrich:2009da} to compute the production cross-sections and model the hard-scatter process, assuming zero angular momentum.
Events were then interfaced to \PYTHIA[8]~\cite{Bierlich:2022pfr} to model parton showering and hadronisation.
As in previous iterations of searches for QBHs, the CTEQ6L1~\cite{Pumplin:2002vw} PDF set was used and the QCD factorisation scale set to the inverse gravitational radius~\cite{Gingrich:2009da}.
Requiring the QBH mass to be in the range $(1{-}3)M_{\textrm{th}}$ (with $M_{\textrm{th}} = M_{\textrm{D}}$) ensures that QBHs are produced in a region where quantum effects are significant, excluding thermal decays.
In the ADD model, $n = 2, 4,$ or $6$ EDs are assumed, whereas $n = 1$ is used in the RS model.
The kinematic distributions of the decay products are governed by the QBH mass; consequently, the signal acceptance and reconstruction efficiency are found to be independent of the specific model type or the number of extra dimensions.
The production cross-section is significantly lower for the RS model than for the ADD model. %
Samples were produced in $0.5~\textrm{TeV}$ steps through threshold mass ranges of $M_{\textrm{th}}\in [6, 8.5]~\textrm{TeV}$ and $[8, 10.5]~\textrm{TeV}$ for the RS and ADD models, respectively.%

Background events from $W$+\;\!jets, $Z$\;\!+\;\!jets and $VV~(WW,~ZZ,~WZ)$ production were modelled with the \SHERPA[2.2.14] generator~\cite{Bothmann:2019yzt} using the \NNPDF[3.0nnlo] PDF set~\cite{Ball:2014uwa}.
The $t\bar{t}$ and single-top processes were modelled using the \POWHEGBOX[v2]~\cite{Frixione:2007nw,Nason:2004rx,Frixione:2007vw, Alioli:2010xd} generator at NLO with the \NNPDF[3.0nlo]~\cite{Ball:2014uwa} PDF set.
The $t\bar{t}W$ process was modelled using \SHERPA[2.2.14], while other $t\bar{t}V$ samples were produced with \MGNLO[2.3.3]~\cite{Alwall:2014hca} interfaced to \PYTHIA[8]~\cite{Sjostrand:2014zea}.
Dijet MC samples were produced with \PYTHIA[8]~\cite{Bierlich:2022pfr}.
\EVTGEN[2.1.1]~\cite{Lange:2001uf} was used to perform bottom- and charm-quark decays except for the backgrounds modelled using \SHERPA, for which the decays were performed internally.
Non-prompt electrons are removed from all MC samples to avoid double-counting their contribution, which is obtained from data.
For the muon channel, this background, estimated from dijet MC events, is negligible.
The same dijet MC sample is used in the electron channel to cross-check the data-driven fake-background estimate.
All generated MC events undergo the full ATLAS detector simulation~\cite{SOFT-2010-01} using \GEANT~\cite{Agostinelli:2002hh}.
The fully simulated MC events are reconstructed in the same manner as the data events and corrected to match the number of interactions per bunch crossing (in-time \pileup) observed in data.
The effect of multiple interactions in the same and neighbouring bunch crossings (\pileup) was modelled by overlaying~\cite{SIMU-2020-01} the simulated hard-scattering event
with inelastic proton--proton events generated from a mix of \EPOS[2.0.1.4]~\cite{Werner:2005jf} and \PYTHIA[8.308]~\cite{Bierlich:2022pfr}.


%

%
\section{Event selection and background estimation}
\label{sec:selection_bkg_estimation}

The physics objects used to reconstruct the lepton+jet final states in this analysis include electrons, muons, jets, and missing transverse momentum.
The definitions of these objects as well as the event selection criteria are given below.

Signal electron candidates are reconstructed from energy clusters in the electromagnetic (EM) calorimeter that are matched to an inner-detector track.
They are required to have $|\eta|<2.47$, excluding the transition region $1.37<|\eta|< 1.52$ between the barrel and endcap EM calorimeters.
To ensure that these electrons originate from the primary vertex, the longitudinal impact parameter $z_0$ of their track relative to the primary vertex must satisfy $|z_0\sin\theta |<0.5~\textrm{mm}$, where $\theta$ is the track's polar angle.
Similarly, their transverse impact parameter $d_0$ must have a significance satisfying $|d_{0}|/\sigma(d_{0})\leq 5$, where $\sigma(d_{0})$ is the uncertainty in $d_{0}$.
Electrons are required to have $p_{\textrm{T}} > 150$~GeV and satisfy the \textit{Tight} identification and \textit{HighPtCaloOnly} isolation criteria~\cite{EGAM-2021-01,EGAM-2021-02}.
They must also match the corresponding trigger-level electron object.

Signal muon candidates are reconstructed by matching inner-detector tracks in the region $|\eta|<2.5$ to tracks reconstructed in the MS.
For muons, the transverse impact parameter requirement is tightened to $|d_{0}|/\sigma(d_{0})\leq 3$, whereas the longitudinal one is the same as for electrons.
Muons must satisfy $p_{\textrm{T}} > 150$~GeV and also the \textit{High}$\text{-}p_{T}$ identification criterion and \textit{PflowTightVarRad} isolation requirement~\cite{MUON-2018-03}.
This identification criterion is used to ensure optimal muon momentum resolution at high $p_{\textrm{T}}$.
It requires at least three hits in each of three layers of precision tracking chambers in the MS, while specific regions of the MS where the alignment is suboptimal are vetoed.
Selected muons are also required to match the trigger-level muon objects.

Jets are reconstructed from particle-flow objects~\cite{PERF-2015-09} using the anti-$k_{t}$ algorithm~\cite{Cacciari:2008gp,Fastjet} with a radius parameter of $R=0.4$.
All jets are required to have $|\eta|<2.5$.
Signal jets must satisfy $p_{\textrm{T}}>130~\textrm{GeV}$.
To resolve ambiguities between reconstructed objects, a standard overlap-removal procedure is applied to leptons and jets to avoid the double-counting of energy deposits~\cite{ATLAS:2025upm}.

The missing transverse momentum, $p^{\textrm{miss}}_{\textrm{T}}$, is defined as the magnitude of the negative vectorial sum of the transverse momenta of all selected and calibrated physics objects and other inner-detector tracks from the primary vertex that are not associated with any of these objects~\cite{JETM-2020-03}.
For the $p^{\textrm{miss}}_{\textrm{T}}$ calculation, the identification criteria for electrons and muons are relaxed. %
Additionally, an object-based missing transverse momentum significance $\mathcal{S}(p^{\textrm{miss}}_{\textrm{T}})$ measures the likelihood that the reconstructed $p^{\textrm{miss}}_{\textrm{T}}$ originates from real invisible particles rather than detector effects.
It is defined as $\mathcal{S}(p^{\textrm{miss}}_{\textrm{T}}) = p^{\textrm{miss}}_{\textrm{T}}/\sqrt{\sigma^{2}_{\textrm{L}}(1 - \rho^{2}_{\textrm{LT}})}$, where $\sigma_{\textrm{L}}$ is the resolution longitudinally to $p^{\textrm{miss}}_{\textrm{T}}$ and $\rho_{\textrm{LT}}$ the correlation between the transverse and longitudinal resolutions relative to $p^{\textrm{miss}}_{\textrm{T}}$~\cite{JETM-2020-03}.
To suppress contributions from \pileup, a jet vertex-tagging~\cite{PERF-2014-03} requirement is applied to jets with $p_{\textrm{T}}<60~\textrm{GeV}$, which may still enter the analysis via the overlap removal or the calculation of $p^{\textrm{miss}}_{\textrm{T}}$.

All reconstructed objects in simulations have their reconstruction efficiency, energy scale and energy resolution corrected to data, where the corrections discussed in Refs.~\cite{EGAM-2021-01,EGAM-2021-02,MUON-2018-03,PERF-2015-09,PERF-2014-03,JETM-2020-03} are updated to the Run~3 conditions.
Moreover, all events are required to contain a primary vertex built from at least two associated tracks with $p_{\textrm{T}} > 0.5~\textrm{GeV}$.
If more than one such vertex is present, the primary vertex is the one with the highest sum of squared transverse momenta $\Sigma p^{2}_{\textrm{T}}$ of its associated tracks~\cite{ATL-PHYS-PUB-2015-026}.

Candidate events in the signal region (SR) require exactly one signal lepton and at least one signal jet with invariant mass $m_{\ell j}>3~\textrm{TeV}$.
The angular separation between the lepton and the jet is required to satisfy $\Delta \phi_{\ell j} > 2.8$ and $\Delta \eta_{\ell j} < 3.25$ to exploit the back-to-back topology of the QBH decay.
To estimate the shape and/or normalisation of the dominant backgrounds, the analysis defines several control and validation regions (CRs and VRs, respectively).
The normalisation of contributions from $W$ and $Z$ events is estimated in single-bin CRs ($W$CR and $Z$CR), while their shapes are taken from MC simulation, following confirmation that no shape correction from data is needed.
These estimates are validated in the respective validation regions ($W$VR and $Z$VR), by examining the ratios of the total background to data and checking that they are within the uncertainty.
Smaller backgrounds from $t\bar{t}$, single-top, $VV$ and $t\bar{t}V$ events are estimated directly from MC simulation.
The background due to fake/non-prompt electrons is estimated from data by using the matrix method (MM)~\cite{ATLAS:2022swp}; the rates determined in a dedicated fake-enriched CR ($f$CR) are applied to all other analysis regions to obtain the fake-electron background contribution there.
The behaviour of this background is examined in a corresponding VR ($f$VR).
In the muon channel, the corresponding contribution is estimated from the dijet MC sample and found to be negligible.
All CRs and VRs are orthogonal to the SRs and to each other via selections on the $m_{\ell j}$ and $\mathcal{S}(p^{\textrm{miss}}_{\textrm{T}})$ variables.
The region definitions and event selection are summarised in Table~\ref{tab:selection}.
The final background estimate is obtained from a simultaneous fit to data in the aforementioned regions.
The fit procedure is described in Section~\ref{sec:result}.

\begin{table}[h!]
\centering
\caption{Analysis regions definitions. The control (validation) regions used to estimate the leading backgrounds, fake electrons and $V$+jets, are indicated. The label $ej (\mu j)$ refers to the electron (muon) channel. The ``signal'' subscript denotes the different lepton and jet signal definitions in the text. The dash (---) indicates a selection criterion not applied, where N/A indicates a criterion cannot be applied, in regions which require exactly one signal lepton. Selections on $m_{\ell j}$ and $\mathcal{S}(p^{\text{miss}}_{\textrm{T}})$ orthogonalise analysis regions from one another. The $f$CR and $f$VR are only relevant in the context of the electron channel. The $m_{\ell\ell}$ criterion in the $f$CR is only applied in the presence of a second electron.}
\begin{tabular}{c c c c | c}
\hhline{=====}
\multirow{2}{*}{Variable}         & \multirow{2}{*}{$W$CR ($W$VR)} & $f$CR ($f$VR)                 & \multirow{2}{*}{$Z$CR ($Z$VR)} & \multirow{2}{*}{SR}      \\
&                                & \small{(electron channel only)} &                                &                          \\ \hline
$N^{j}_{\textrm{signal}}$         & $\geq 1$                       & $\geq 1$                       & $\geq 1$                        & $\geq 1$                 \\ \Tstrut\Bstrut
$N^{\ell}_{\textrm{signal}}$      & 1                              & 1                              & 2                               & 1                        \\ \Tstrut\Bstrut
$m_{\ell j}$ {[}TeV{]}            & 1.0--2.0 (2.0--3.0)            & 1.0--3.0                       & 1.0--2.0 (2.0--3.0)             & \textgreater 3.0         \\ \Tstrut\Bstrut
$p^{\ell}_{\textrm{T}}$ {[}GeV{]} & \textgreater 150               & \textgreater{}150              & \textgreater 150                & \textgreater 150         \\ \Tstrut\Bstrut
$p^{j}_{\textrm{T}}$ {[}GeV{]}    & \textgreater 130               & \textgreater 130               & \textgreater 130                & \textgreater 130         \\ \Tstrut\Bstrut
$\mathcal{S}(p^{\text{miss}}_{\textrm{T}})~ej, \mu{j}$ & \textgreater 5.0, 3.0 & \textless 3.0 (3.0--5.0) & ---                             & ---                      \\ \Tstrut\Bstrut
$m_{\ell\ell}$ {[}GeV{]}          & N/A                            & \textgreater 120               & 60--120                         & N/A                      \\ \Tstrut\Bstrut
$\Delta\eta_{\ell j}$             & \textless 3.25                 & ---                            & ---                             & \textless 3.25           \\ \Tstrut\Bstrut
$\Delta\phi_{\ell j}$             & \textgreater 2.8               & ---                            & ---                             & \textgreater 2.8         \\ \hhline{=====}
\end{tabular}
\label{tab:selection}
\end{table}

\begin{figure}[H]
\subfloat[]{
\includegraphics[width=0.5\textwidth]{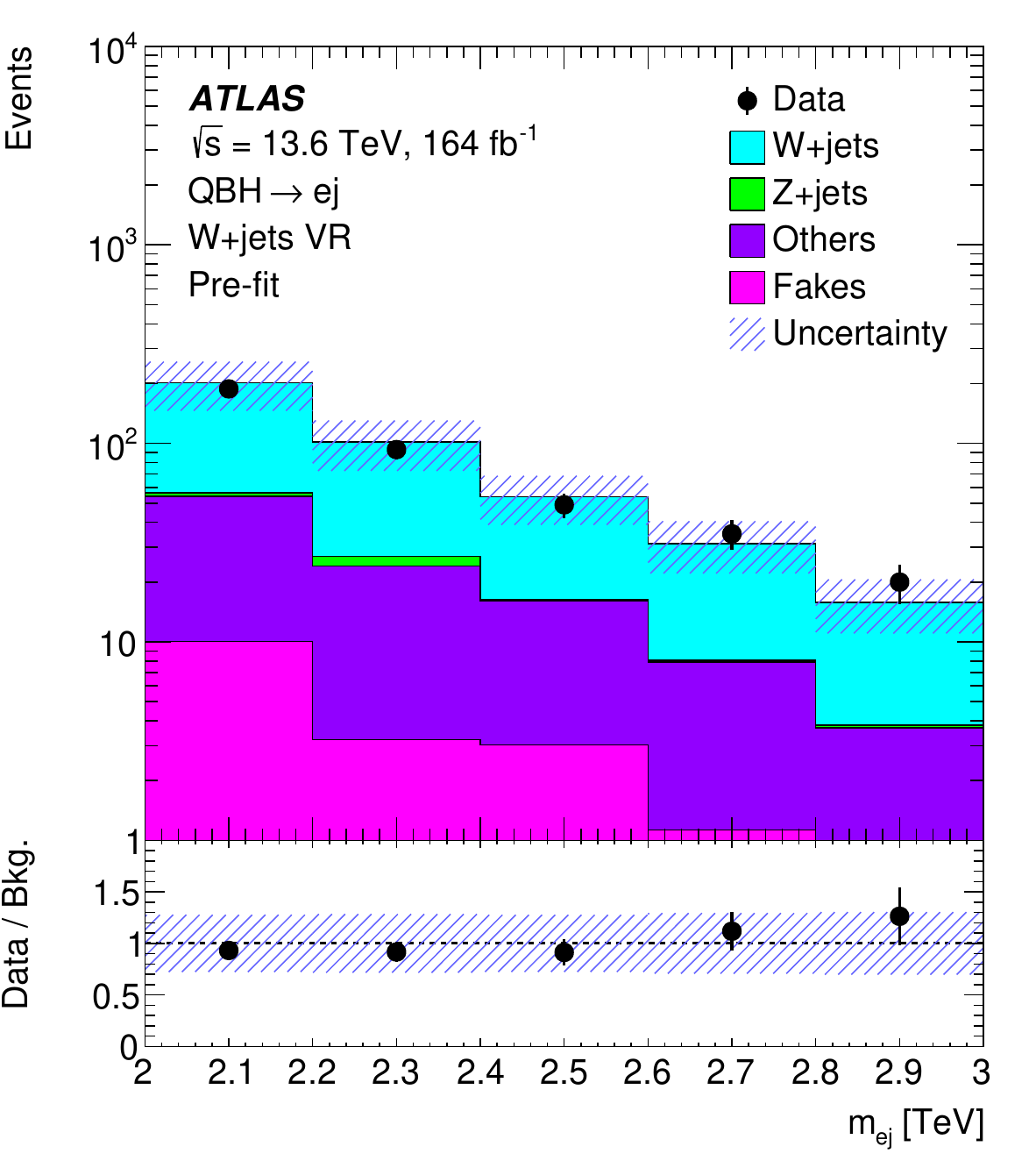}
}
\subfloat[]{
\includegraphics[width=0.5\textwidth]{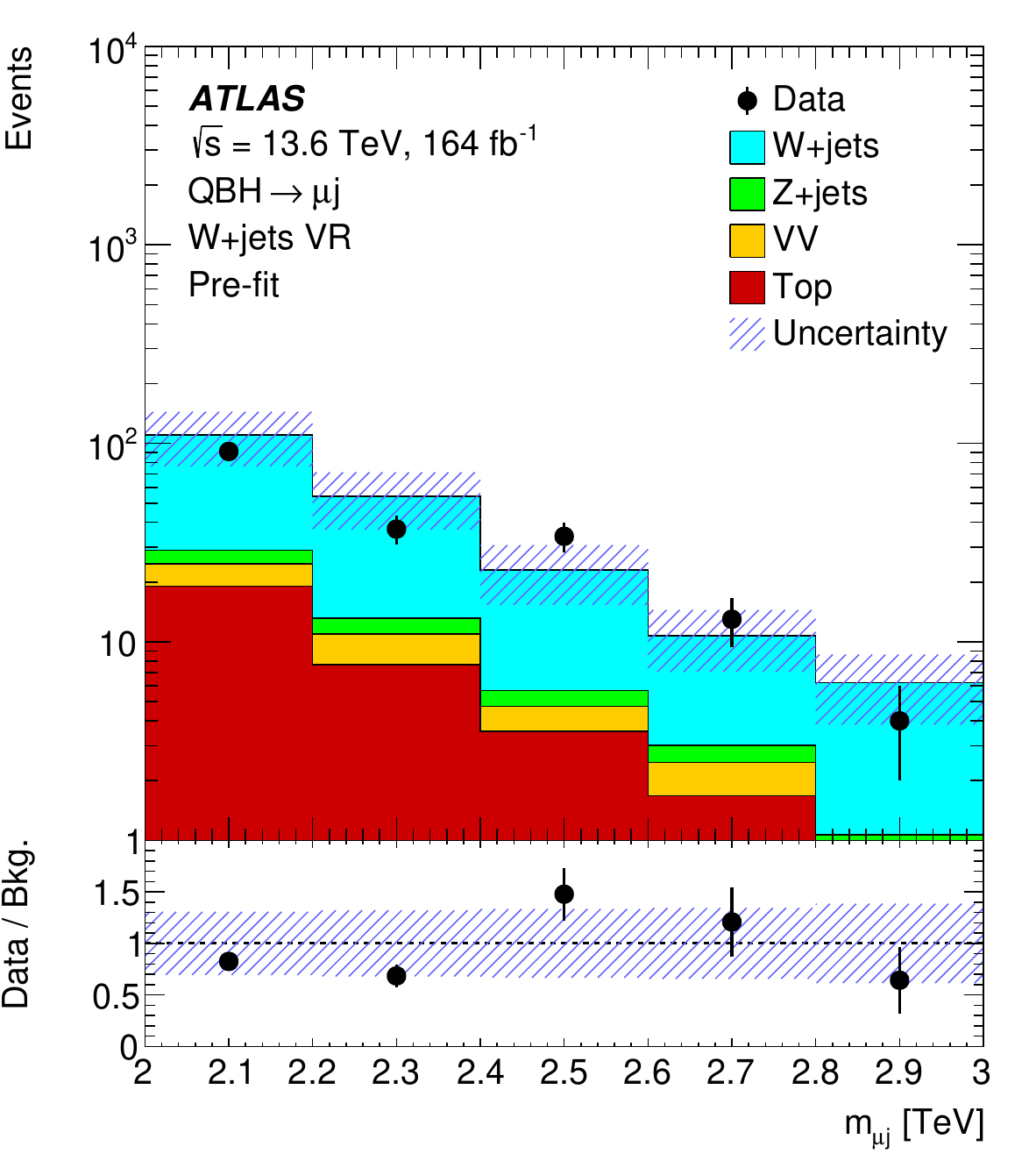}
}
\caption{Invariant mass of the lepton--jet pair in the $W$+\;\!jets validation region for (a) the electron channel and (b) the muon channel. The data-to-background ratio is shown in the bottom panels, and hatched bands indicate the total statistical and systematic uncertainty. In (a), \enquote{Others} denotes subdominant backgrounds including top and diboson processes. The MC contributions are normalised based on their expected cross-section, prior to the likelihood fit.}
\label{fig:Wjets_VR_mLepJet_ManyBins}
\end{figure}

The $W$CR is designed to be maximally pure in the $W$+\;\!jets background component by requiring $\mathcal{S}(p^{\textrm{miss}}_{\textrm{T}}) >5.0\,(3.0)$ for the electron (muon) channel.
The different selection values originate from the need to accommodate the $f$CR in the electron channel, as discussed below.
Similarly, the $Z$CR is made maximally pure in the $Z$\;\!+\;\!jets component by requiring a second same-flavour lepton, with the dilepton invariant mass constrained to the range $60$--$120~\textrm{GeV}$. These two CRs are defined in the $m_{\ell j}$ range from $1$ to $2~\textrm{TeV}$, while the respective VRs cover the $2$ to $3~\textrm{TeV}$ range.
Distributions in the $W$VRs are shown in Figure~\ref{fig:Wjets_VR_mLepJet_ManyBins}, where it is shown that data and background predictions agree well within the assigned uncertainties.

\begin{figure}[H]
\centering
\includegraphics[width=0.6\textwidth]{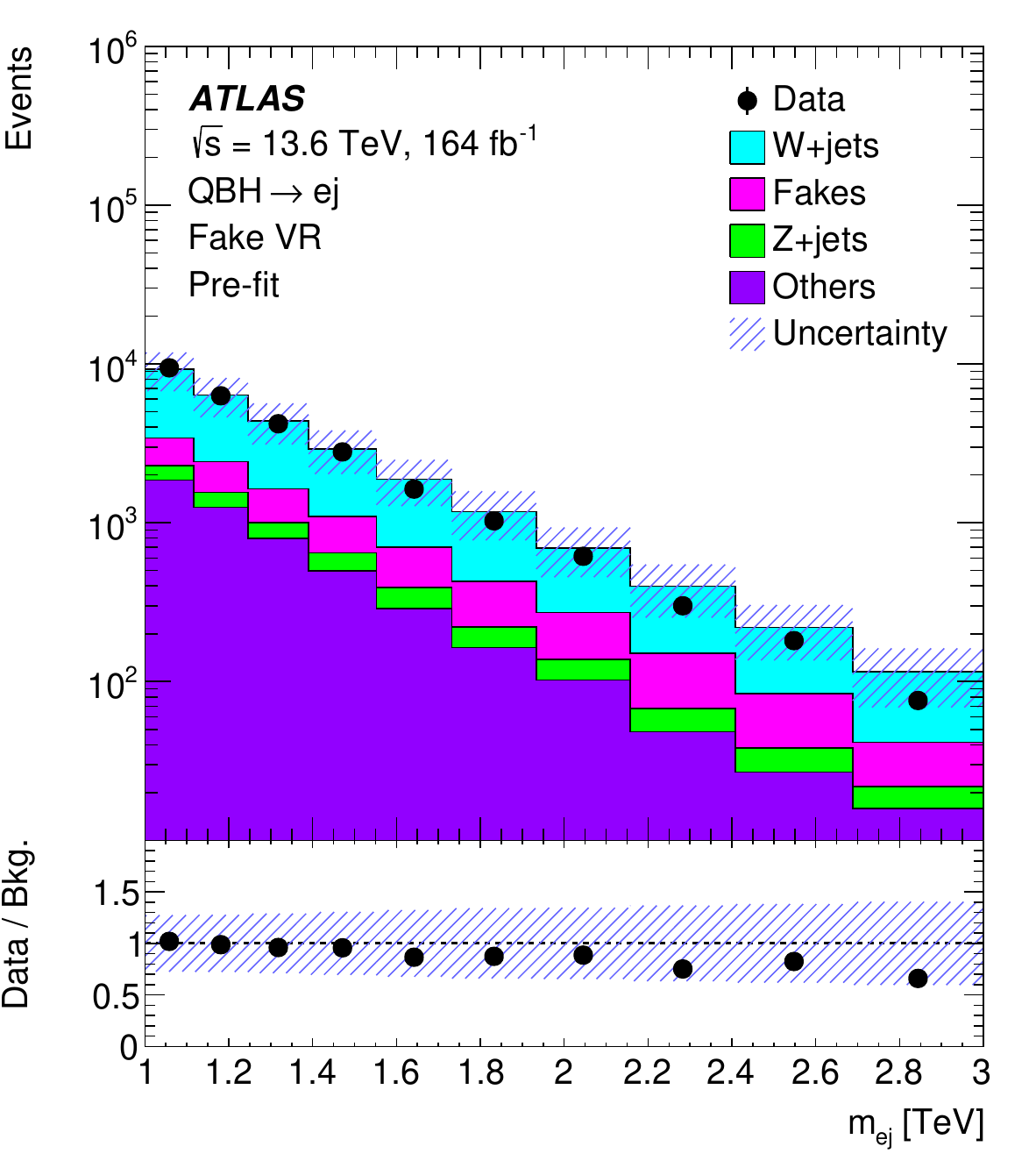}
\caption{Invariant mass of the lepton--jet pair in the fakes validation region. The data-to-background ratio is shown in the bottom panel, and hatched bands indicate the total statistical and systematic uncertainty. \enquote{Others} denotes subdominant backgrounds including top and diboson processes. The MC contributions are normalised based on their expected cross-section, prior to the likelihood fit.}
\label{fig:fake_VR}
\end{figure}

In the electron channel, the dominant background in the SR arises from fakes, estimated via the MM.
The MM requires the definition of \enquote{real} and \enquote{fake} rates, $r$ and $f$, respectively.
These probabilities are defined as the ratio of events coming from two different electron populations: signal and baseline.
The signal criteria are defined above.
The baseline criteria are defined by relaxing the electron identification from \textit{Tight} to \textit{Medium} and removing the isolation requirement~\cite{EGAM-2021-01,EGAM-2021-02}.
These criteria are only used in order to (i) estimate $f$ in $f$CR, and (ii) build the respective background template in the different regions.
The estimation of $r$ is performed in a \enquote{real}-enriched region using $Z(\rightarrow e^{+}e^{-})\;\!+\;\!$jets MC events.
The estimation of $f$ is performed in a \enquote{fake}-enriched region that is both orthogonal to the SR and contains a high-enough event yield.
This is achieved by requiring that $m_{\ell j}$ be in the range $1$--$3~\textrm{TeV}$.
To increase the fakes' purity in that region, contributions from the dominant $W$+\;\!jets and $Z$\;\!+\;\!jets backgrounds are reduced by placing a requirement on $\mathcal{S}(p^{\textrm{miss}}_{\textrm{T}})$ or $m_{ee}$, respectively.
The resulting fraction of events with fakes in the $f$CR ranges from $60\%$ to $90\%$ ($30\%$ to $45\%$) for the baseline (signal) population.
Both $r$ and $f$ are calculated as a function of the electron $p_{\textrm{T}}$ in the range 150--2000~GeV and $|\eta|$ in four bins within the range 0--2.47, excluding the transition region mentioned above.
In the calculation of $f$, the prompt-electron contributions found in MC simulation are subtracted from the data in the $f$CR, for both populations separately.
Using the MM, $r$ and $f$ are combined into weights that are applied to data events meeting the baseline criteria.
This operation yields the fakes background template in the different regions.
The result is validated in the $f$VR, defined in Table~\ref{tab:selection}.
The $m_{\ell j}$ distribution in the $f$VR is shown in Figure~\ref{fig:fake_VR}, where the data and background predictions are seen to be in agreement within the assigned uncertainties.
As can be seen, the fakes component is subdominant in this region, roughly 2--3 times smaller than the $W$+\;\!jets component.
Nevertheless, this still allows to assess the modelling of the fakes background estimation there.

In the electron channel prior to the fit, the fakes component constitutes approximately $56\%$ of the total SR background, followed by $W$+\;\!jets at $32\%$, with $Z$\;\!+\;\!jets, top, and diboson processes contributing the remaining $12\%$.
In the muon channel, $W$+\;\!jets dominates at approximately $80\%$, with $Z$\;\!+\;\!jets, top, and diboson processes accounting for the remaining $20\%$.
The signal acceptance times efficiency ranges from $76\%$ to $86\%$ ($52\%$ to $58\%$) for the electron (muon) channel, for the signals considered in this analysis.
The factors responsible for the difference in yield between the two channels are the lepton acceptance times efficiency and the lepton momentum resolution.
The muon trigger has a lower geometric coverage than the electron counterpart.
Furthermore, very tight identification requirements are necessary to maintain a reasonable muon momentum resolution at high $p_{\textrm{T}}$.
Uncertainties associated with the modelling of high-$p_{\textrm{T}}$ muon identification are taken into account as discussed in Ref.~\cite{MUON-2018-03}.
The limited momentum resolution leads to a significantly worse muon+jet mass resolution as compared to the electron+jet case and thus broader signal shapes.
The lepton+jet mass resolution in the electron channel is consistently below 5\% for all signal $M_{\textrm{th}}$ points considered in the analysis.
The muon-channel mass resolution varies between 25\% and 30\% for signal $M_{\textrm{th}}$ values between $6~\textrm{TeV}$ and $10~\textrm{TeV}$, respectively.
The resolution is obtained from the quadrature sum of the standard deviation and the small bias in the mean of the respective lepton+jet mass reconstruction response distribution.


%

%
\section{Systematic uncertainties}
\label{sec:systematics}

Systematic uncertainties affecting the signal and background expectations arise from various experimental and theoretical sources and are evaluated using the methods described below.
Theoretical uncertainties are only calculated for the two leading MC backgrounds, $W$+\;\!jets and $Z$+\;\!jets, since the others are significantly smaller and their uncertainties have negligible impact on the analysis.
These uncertainties, which originate from missing higher-order corrections, contribute via the choice of QCD renormalisation and factorisation scales, strong coupling constant and PDFs~\cite{Ball:2014uwa}.
These sources are collectively referred to as \enquote{theory uncertainties} and amount to ${\sim}30\%$ of the $V$+\;\!jets background yield in the SR.
Following previous publications, no theoretical uncertainties are assigned to the QBH benchmark signals.
This choice is related to the unclarity in the application of these uncertainties to QBH production.
It originates from the lack of an established quantum theory for the interaction between gravity and partons.
Experimental uncertainties are evaluated for all MC signal and background samples.
These uncertainties include those estimated by varying the reconstruction, identification, isolation and trigger efficiencies, and the energy/momentum calibration of electrons~\cite{EGAM-2018-01}, muons~\cite{MUON-2018-03}, jets~\cite{JETM-2018-05} and $p^{\textrm{miss}}_{\textrm{T}}$~\cite{JETM-2020-03}, as well as uncertainties in the integrated luminosity~\cite{ATL-DAPR-PUB-2025-001} and \pileup.
The theory and experimental uncertainties are propagated to the fakes background through the prompt-electron MC event yield that is subtracted from data in the $f$CR, as explained in the previous section.
Collectively, they correspond to an uncertainty of approximately 18\% of the fakes background in the SR, with the dominant contribution being from QCD scale uncertainties.
Additionally, a separate normalisation uncertainty due to the fakes in the $f$CR and SR not having the same mix of origins is considered.
In this analysis, almost all fakes originate from light-flavour decays or photon conversions, according to the dijet MC sample.
The two origins populate the $f$CR and SR differently.
Light-flavour decays account for 92\% (71\%) of fakes in the $f$CR (SR), whereas photon conversions account for 8\% (29\%).
The corresponding uncertainty is found by calculating the weighted average of the difference in the two origins between the $f$CR and SR, as follows: let $g^{i,j}_{u,v}$ be the fraction of electrons of origin $\{\text{light-flavour decays, photon conversion}\}$ in region $\{f\text{CR, SR}\}$, where LF denotes light-flavour decays and $\gamma$-conv denotes photon conversions, and $g^{\text{LF}}_{\text{SR}}+g^{\gamma\text{-conv}}_{\text{SR}} = 1$.
The uncertainty is therefore $g^{\text{LF}}_{\text{SR}} |g^{\text{LF}}_{f\text{CR}}-g^{\text{LF}}_{\text{SR}}| + g^{\gamma\text{-conv}}_{\text{SR}} |g^{\gamma\text{-conv}}_{f\text{CR}}-g^{\gamma\text{-conv}}_{\text{SR}}|$.
This uncertainty amounts to 22\% on the yield of the fakes background.
By considering the difference in origin between the $f$CR and $f$VR, the same uncertainty, computed in the same manner, is found to be 17\% in the $f$VR.
Lastly, the limited size of the data and MC samples in the $f$CR is accounted for via a separate systematic uncertainty, which is then propagated as a function of electron $p_{\textrm{T}}$ and $|\eta|$ to the SR, where it amounts to about 16\%.
%
%
%


%
%
%
%
%
%
%
%

%
\section{Results}
\label{sec:result}

A signal-plus-background model is used to fit the observed $m_{\ell j}$ spectra simultaneously in the SR, $W$CR and $Z$CR using the profile-likelihood method~\cite{Cowan:2010js}.
The expected and observed event counts in each $m_{\ell j}$ bin are described by Poisson-distributed variables, where the uncertainties are accounted for via nuisance parameters.
The SR bins are chosen such that all have a background expectation of $\geq \mathcal{O}(1)$ events and are coarser than the detector resolution.
The sensitivity of the analysis is driven by the highest $m_{\ell j}$ bin of the SR, 5--13~TeV.
Contributions above 13~TeV are included in this bin.
In that bin, the number of background events in the electron channel is predicted to be \textcolor{black}{$1.46\pm 0.65_{\textrm{stat.}}\pm 0.38_{\textrm{syst.}}$}, where the leading systematic uncertainty is from the fakes estimation.
In the muon channel, the number of background events in the highest bin is predicted to be \textcolor{black}{$3.10\pm 0.95_{\textrm{stat.}}\pm 0.47_{\textrm{syst.}}$}, where the leading systematic uncertainty is due to the modelling of the $W$+\;\!jets background.
Normalisation factors for the leading MC backgrounds, $W$+\;\!jets and $Z$\;\!+\;\!jets, are found to be $1.2\pm 0.5$ and $1.1\pm 0.4$ ($1.1\pm 0.5$ and $1.0\pm 0.3$) for $W$+\;\!jets and $Z$\;\!+\;\!jets in the electron (muon) channel, respectively.
A moderate 20\% (25\%) non-closure is observed in the $Z$VR in the electron (muon) channel; given the small $Z$\;\!+\;\!jets fraction of 5\% (12\%) in the SR, the potential impact on the background yield in the SR is at most 1\%--2\% and is negligible. It is therefore not included in the fit as a separate uncertainty.

\begin{figure}[htbp]
\subfloat[]{
\includegraphics[width=0.5\textwidth]{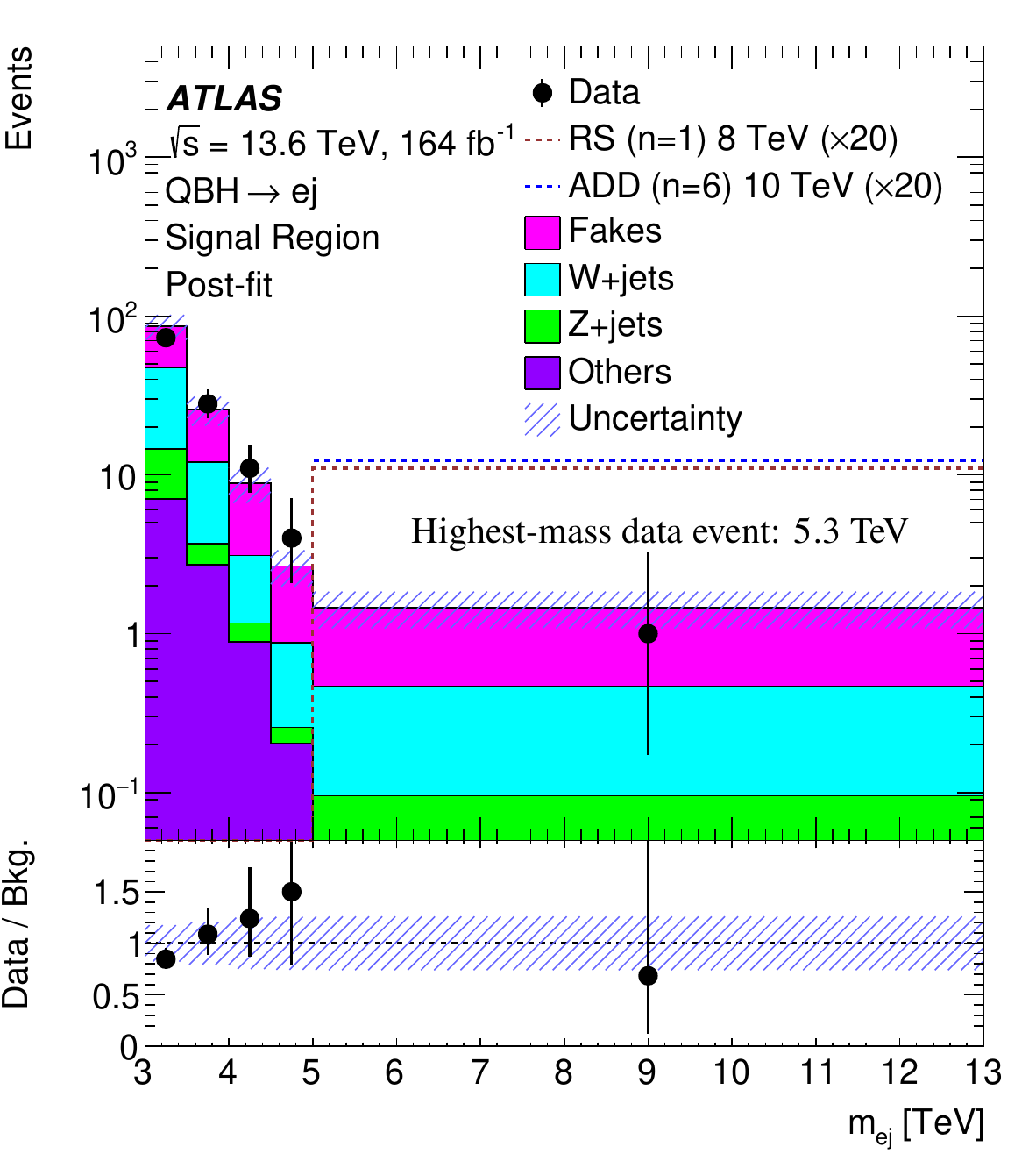}
}
\subfloat[]{
\includegraphics[width=0.5\textwidth]{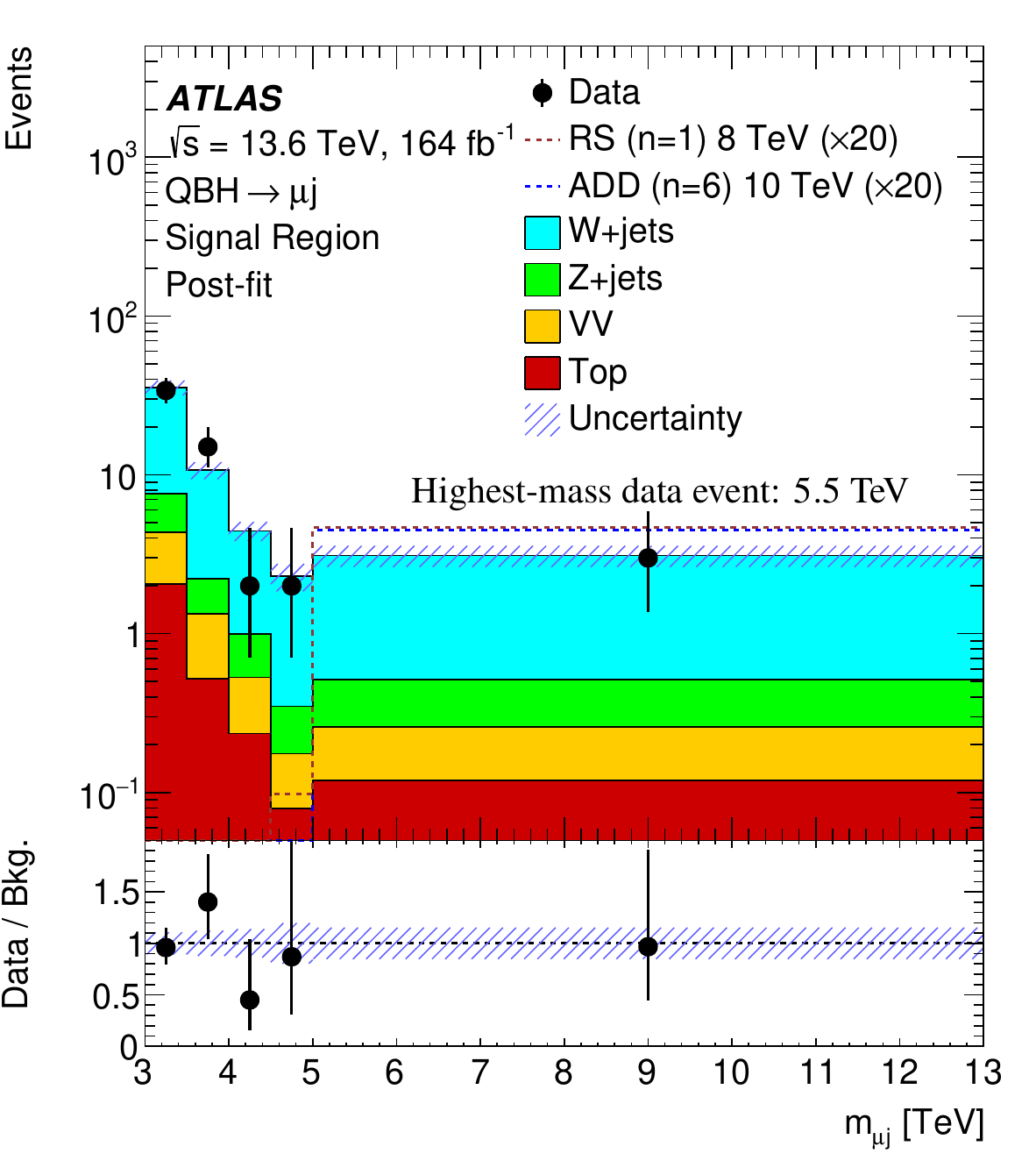}
}
\caption{Distribution of the lepton+jet invariant mass for events passing the full signal selection for (a) the electron channel and (b) the muon channel. Benchmark signals for the RS and ADD models with $n=1,6$ extra dimensions at $M_{\textrm{th}}=8,10~\textrm{TeV}$, respectively, are scaled by a factor of 20 for presentation purposes and overlaid on the total background estimate. The uncertainty band includes the statistical and systematic uncertainties. The highest-mass data events observed in the last bin have $m_{\ell j}$ values of 5.3~(5.0, 5.1, 5.5)~TeV in the electron (muon) channel, and are plotted at the bin's centre. In (a), \enquote{Others} denotes subdominant backgrounds including top and diboson processes. The MC contributions are normalised based on the likelihood fit.}
\label{fig:SR_mLepJet}
\end{figure}

No significant excess of data is observed above the SM background.
The $m_{\ell j}$ distributions in the electron channel SR and muon channel SR after the fit are shown in Figure~\ref{fig:SR_mLepJet}.
Benchmark signal MC predictions corresponding to the RS and ADD models are overlaid for reference.
The signals considered in this search are generally expected to populate the $m_{\ell j}$ region above 6~TeV, while the background shape falls steeply between 3 and 6~TeV in both channels.
In the electron channel, the highest-mass event observed in data corresponds to $m_{\ell j} = 5.3~\textrm{TeV}$.
This event features an electron with $(p_{\textrm{T}}, \eta) = (1.2~\textrm{TeV}, 1.55)$, a jet with $(p_{\textrm{T}}, \eta) = (1.2~\textrm{TeV}, -1.29)$, and a low $\mathcal{S}(p_{\textrm{T}}^{\textrm{miss}})$ of $2.80$.
The objects are nearly back-to-back in the transverse plane with $\Delta\phi(\ell, j) = 3.05$, with a large pseudorapidity separation $\Delta\eta(\ell, j) = 2.84$.
In the muon channel, the event with the highest invariant mass corresponds to $m_{\ell j} = 5.5~\textrm{TeV}$.
This event features a muon with $(p_{\textrm{T}}, \eta) = (5.2~\textrm{TeV}, 1.33)$, a jet with $(p_{\textrm{T}}, \eta) = (1.2~\textrm{TeV}, 0.39)$, and a low $\mathcal{S}(p_{\textrm{T}}^{\textrm{miss}})$ of $1.43$.
The objects are nearly back-to-back in the transverse plane with $\Delta\phi(\ell, j) = 3.11$, with a relatively small pseudorapidity separation of $\Delta\eta(\ell, j) = 0.94$ indicating a centrally produced system.

\begin{figure}
\subfloat[]{
\includegraphics[width=0.5\textwidth]{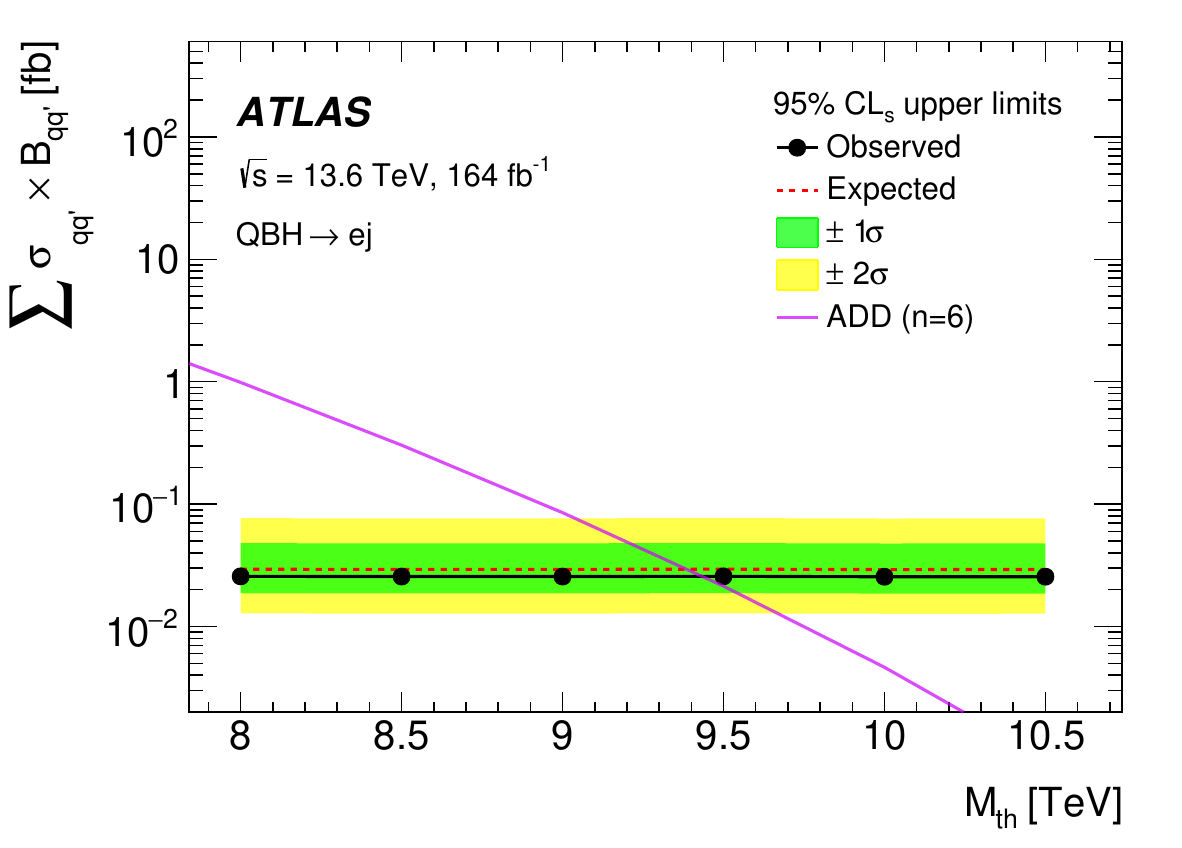}
}
\subfloat[]{
\includegraphics[width=0.5\textwidth]{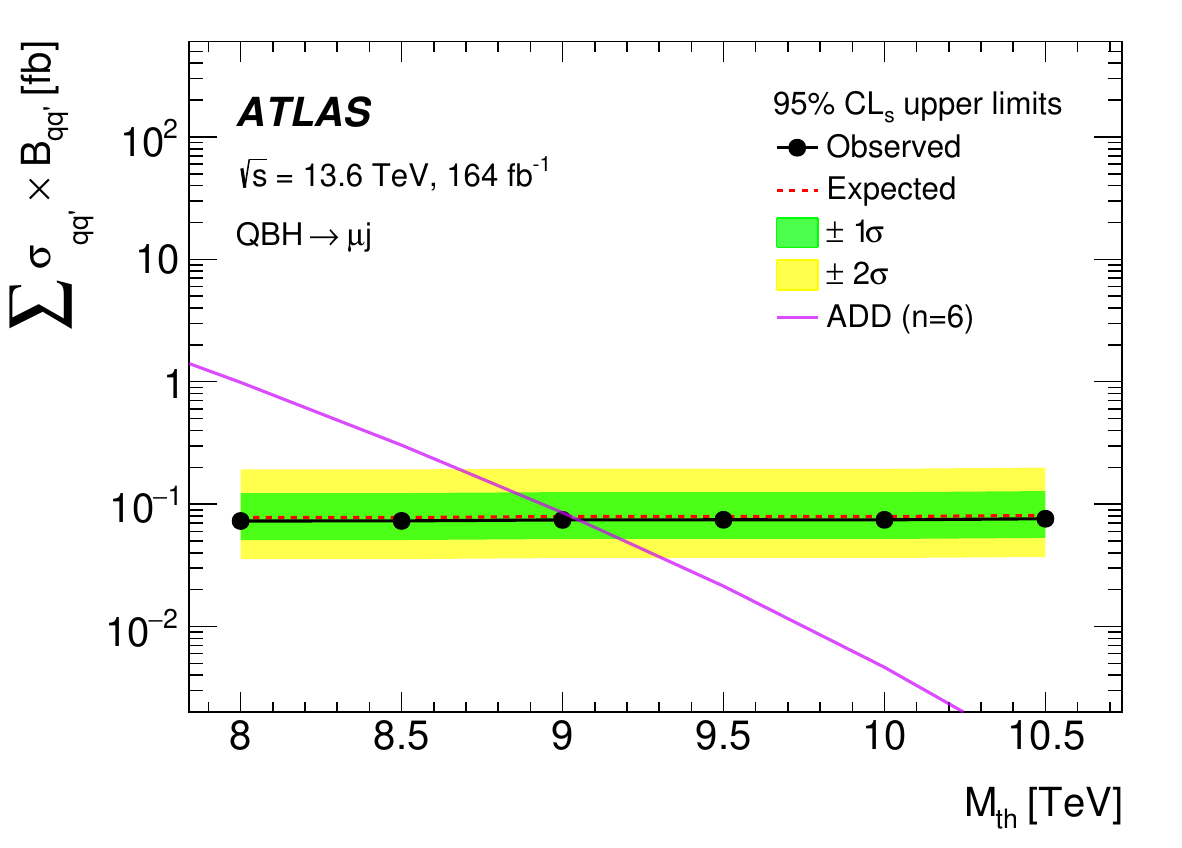}
}\\
\subfloat[]{
\includegraphics[width=0.5\textwidth]{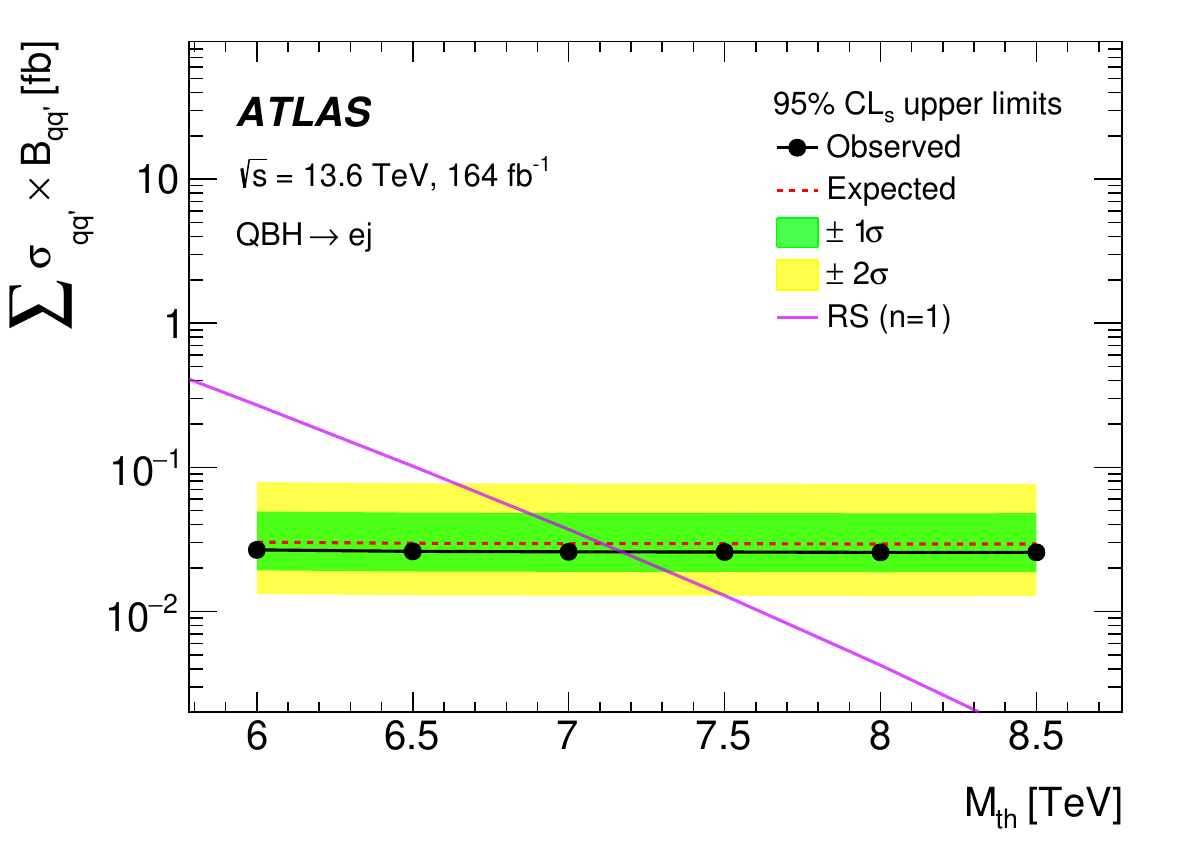}
}
\subfloat[]{
\includegraphics[width=0.5\textwidth]{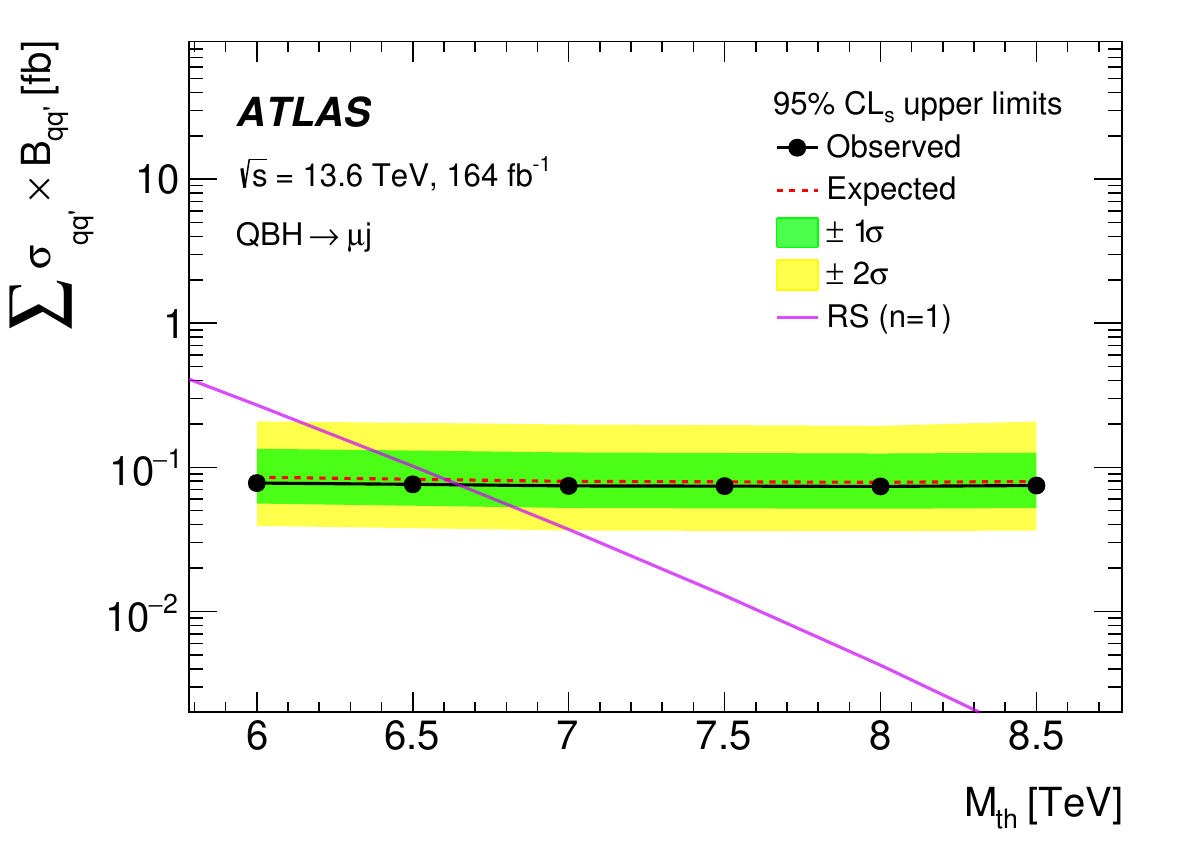}
}
\caption{Exclusion limits at 95\% confidence level (CL) on the sum of the production cross-section times branching ratio of quantum black holes in the ADD model with $n=6$ extra dimensions for the (a) electron and (b) muon channels, and in the RS model with $n=1$ extra dimension for the (c) electron and (d) muon channels. The $\pm 1\sigma$ and $\pm 2\sigma$ uncertainty bands around the expected limit are shown. The central theoretical values are overlaid. The summation subscript $qq'$ corresponds to the three relevant production channels of the quantum black hole. The branching fraction used is per lepton flavour.}
\label{fig:limits}
\end{figure}

Upper limits on the QBH production cross-section times branching ratio are set at 95\% CL using the CL$_\text{s}$ method in the asymptotic approximation, as shown in Figure~\ref{fig:limits}.
The limits set using the asymptotic approximation are validated against those obtained with pseudo-experiments, showing per-mille-level agreement on the cross-section limit for a variety of tested mass hypotheses.

The RS model is excluded at 95\% CL for signals with $M_{\textrm{th}}<7.2~\textrm{TeV}$.
Depending on the number of EDs, ADD models with $M_{\textrm{th}}$ below 8.6--9.4~TeV are excluded at 95\% CL.
The observed and expected $M_{\textrm{th}}$ exclusion limits at 95\% CL for the signal models considered are detailed in Table~\ref{tab:intersection_limits}.
The electron channel surpasses the muon channel in sensitivity by a factor of ${\sim}3$, due to better momentum resolution for electrons and its impact on the acceptance times efficiency, as well as the lower background in the SR.
Hence, the two channels are not combined.

\begin{table}[h!]
\centering
\caption{Expected and observed exclusion limits at 95\% confidence level on the QBH mass threshold $M_{\mathrm{th}}$ for RS and ADD models with different number of extra dimensions $n$, in the electron and muon channels, using Run~3 data at a centre-of-mass energy of 13.6 TeV.}
\begin{tabular}{ll|cc|cc|cc|cc}
\hhline{==========}
Model   & & \multicolumn{2}{c|}{RS} & \multicolumn{6}{c}{ADD} \\
\hhline{----------}
$n$     & & \multicolumn{2}{c|}{1}  & \multicolumn{2}{c|}{2} & \multicolumn{2}{c|}{4} & \multicolumn{2}{c}{6} \\
\hhline{==========}
Channel & & $e+j$ & $\mu+j$ & $e+j$ & $\mu+j$ & $e+j$ & $\mu+j$ & $e+j$ & $\mu+j$ \\
\hhline{==========}
Exclusion limit & & \multicolumn{8}{c}{Expected} \\
\hhline{----------}
$M_{\mathrm{th}}$ [TeV] & & 7.1 & 6.6 & 8.5 & 8.1 & 9.1 & 8.7 & 9.4 & 9.0 \\
\hhline{==========}
Exclusion limit   & & \multicolumn{8}{c}{Observed} \\
\hhline{----------}
$M_{\mathrm{th}}$ [TeV] & & 7.2 & 6.6 & 8.6 & 8.2 & 9.1 & 8.7 & 9.4 & 9.0 \\
\hhline{==========}
\end{tabular}
\label{tab:intersection_limits}
\end{table}



%
%
%
%
%
%
%
%
%
%
%
%
%
%

%
%
\FloatBarrier

\section{Conclusion}
\label{sec:conclusion}

This ATLAS analysis is performed using $164~\textrm{fb}^{-1}$ of data collected from $\sqrt{s}=13.6~\textrm{TeV}$ $pp$ collisions at the LHC.
The search for QBH-like signals yields no evidence of an excess in the lepton+jet invariant mass distribution in electron+jet or muon+jet final states.
Exclusion limits are placed on the production cross-section times branching ratio in a few ADD and RS QBH scenarios. %
The experimental sensitivity is driven by the statistical uncertainty.
The highest exclusion limits are for ADD QBH scenarios with six extra dimensions, observed at 9.4~TeV (9.0~TeV) in the electron (muon) channel.
Owing to the much better momentum resolution and larger acceptance times efficiency in the electron channel, the sensitivity is significantly higher.
This analysis extends the mass-exclusion reach of the previous ATLAS search performed at $\sqrt{s} = 13~\textrm{TeV}$ using the full $140~\textrm{fb}^{-1}$ Run~2 dataset~\cite{ATLAS:2023vat} by $0.8~(0.4)~\textrm{TeV}$ in the electron (muon) channel for QBHs in the ADD model with $n=6$ extra dimensions.
These limits represent the world-best sensitivity for the signals considered.
%
%


%
%
%
%
%
%
%
%

%
%
%
%
%
\section*{Acknowledgements}
%

%

%
%

%
%

We thank CERN for the very successful operation of the LHC and its injectors, as well as the support staff at
CERN and at our institutions worldwide without whom ATLAS could not be operated efficiently.

The crucial computing support from all WLCG partners is acknowledged gratefully, in particular from CERN, the ATLAS Tier-1 facilities at TRIUMF/SFU (Canada), NDGF (Denmark, Norway, Sweden), CC-IN2P3 (France), KIT/GridKA (Germany), INFN-CNAF (Italy), NL-T1 (Netherlands), PIC (Spain), RAL (UK) and BNL (USA), the Tier-2 facilities worldwide and large non-WLCG resource providers. Major contributors of computing resources are listed in Ref.~\cite{ATL-SOFT-PUB-2026-001}.

We gratefully acknowledge the support of ANPCyT, Argentina; YerPhI, Armenia; ARC, Australia; BMWFW and FWF, Austria; ANAS, Azerbaijan; CNPq and FAPESP, Brazil; NSERC, NRC and CFI, Canada; CERN; ANID, Chile; CAS, MOST, NSFC and FRFCU, China; Minciencias, Colombia; MEYS CR, Czech Republic; DNRF and DNSRC, Denmark; IN2P3-CNRS and CEA-DRF/IRFU, France; SRNSFG, Georgia; BMFTR, HGF and MPG, Germany; GSRI, Greece; RGC and Hong Kong SAR, China; ICHEP and Academy of Sciences and Humanities, Israel; INFN, Italy; MEXT and JSPS, Japan; CNRST, Morocco; NWO, Netherlands; RCN, Norway; MNiSW, Poland; FCT, Portugal; MNE/IFA, Romania; MSTDI, Serbia; MSSR, Slovakia; ARIS and MVZI, Slovenia; DSI/NRF, South Africa; MICIU/AEI, Spain; SRC and Wallenberg Foundation, Sweden; SERI, SNSF and Cantons of Bern and Geneva, Switzerland; NSTC, Taipei; TENMAK, T\"urkiye; STFC/UKRI, United Kingdom; DOE and NSF, United States of America.

Individual groups and members have received support from BCKDF, CANARIE, CRC and DRAC, Canada; CERN-CZ, FORTE and PRIMUS, Czech Republic; COST, ERC, ERDF, Horizon 2020 and Marie Sk{\l}odowska-Curie Actions, European Union; Investissements d'Avenir Labex, Investissements d'Avenir Idex and ANR, France; DFG and AvH Foundation, Germany; Herakleitos, Thales and Aristeia programmes co-financed by EU-ESF and the Greek NSRF, Greece; BSF-NSF and MINERVA, Israel; NCN and NAWA, Poland; La Caixa Banking Foundation, CERCA and AGAUR programs from Generalitat de Catalunya and PROMETEO and GenT Programmes Generalitat Valenciana, Spain; G\"{o}ran Gustafssons Stiftelse, Sweden; The Royal Society and Leverhulme Trust, United Kingdom; Eric and Wendy Schmidt Fund for Strategic Innovation, United States of America.

In addition, individual members wish to acknowledge support from Chile: Agencia Nacional de Investigaci\'on y Desarrollo (ANID FONDECYT reg. 1230987, FONDECYT 1230812, FONDECYT 1240864, Fondecyt 3240661, Fondecyt Regular 1240721); China: Fundamental Research Funds for the Central Universities (010-63263105), Chinese Ministry of Science and Technology (MOST-2023YFA1605700, MOST-2023YFA1609300), National Natural Science Foundation of China (NSFC 12275265, NSFC-W2543005); Czech Republic: Czech Science Foundation (GACR - 24-11373S), Ministry of Education Youth and Sports (ERC-CZ-LL2327, FORTE CZ.02.01.01/00/22\_008/0004632), PRIMUS Research Programme (PRIMUS/21/SCI/017); EU: H2020 European Research Council (ERC - 101002463); European Union: European Research Council (BARD No. 101116429, ERC - 101219398, ERC - 948254, ERC 101089007), European Regional Development Fund (HE COFUND GA No.101081355, ERDF), Marie Sklodowska-Curie Actions (GAP-101168829); France: Agence Nationale de la Recherche (ANR-21-CE31-0013, ANR-22-EDIR-0002, ANR-24-CE31-0504-01); Germany: Deutsche Forschungsgemeinschaft (DFG - 469666862); China: Research Grants Council (GRF); Italy: Istituto Nazionale di Fisica Nucleare (LHC-MIUR - 28003/2025), Ministero dell'Università e della Ricerca (NextGenEU 153D23001490006 M4C2.1.1, NextGenEU I53D23000820006 M4C2.1.1, NextGenEU I53D23001490006 M4C2.1.1, SOE2024\_0000023); Japan: Japan Society for the Promotion of Science (JSPS KAKENHI  JP25H0063, JSPS KAKENHI JP22H01227, JSPS KAKENHI JP22H04944, JSPS KAKENHI JP22KK0227, JSPS KAKENHI JP24K23939, JSPS KAKENHI JP24KK0251, JSPS KAKENHI JP25H00650, JSPS KAKENHI JP25H01291, JSPS KAKENHI JP25K01011, JSPS KAKENHI JP25K01023); Poland: Polish National Science Centre (NCN 2021/42/E/ST2/00350, NCN OPUS 2023/51/B/ST2/02507, NCN OPUS nr 2022/47/B/ST2/03059, NCN UMO-2019/34/E/ST2/00393, UMO-2022/47/O/ST2/00148, UMO-2023/49/B/ST2/04085, UMO-2023/51/B/ST2/00920, UMO-2024/53/N/ST2/00869); Spain: Agència de Gestió d'Ajuts Universitaris i de Recerca. (AGAUR - 2023 BP 00141), Ministry of Science and Innovation (RYC2019-028510-I, RYC2020-030254-I, RYC2021-031273-I, RYC2022-038164-I), Ministerio de Ciencia, Innovación y Universidades/Agencia Estatal de Investigaci\'on (EU NextGenerationEU (PRTR-C17.I1), PID2022-142604OB-C22); Sweden: Carl Trygger Foundation (Carl Trygger Foundation CTS 22:2312), Swedish Research Council (Swedish Research Council 2023-04654, VR 2021-03651, VR 2022-03845, VR 2022-04683, VR 2023-03403, VR 2024-05451, VR 2025-05940), Knut and Alice Wallenberg Foundation (KAW 2023.0366); Switzerland: Swiss National Science Foundation (SNSF - PCEFP2\_194658); United Kingdom: The Binks Trust, Royal Society (NIF-R1-231091); United States of America: U.S. Department of Energy (ECA DE-AC02-76SF00515), John Templeton Foundation (John Templeton Foundation 63206), Neubauer Family Foundation.

%
%


%
\section*{Data availability}
\printatlasdatapolicy
\clearpage
\printbibliography
\clearpage
 
\begin{flushleft}
\hypersetup{urlcolor=black}
{\Large The ATLAS Collaboration}

\bigskip

\AtlasOrcid[0000-0002-6665-4934]{G.~Aad}$^\textrm{\scriptsize 102}$,
\AtlasOrcid[0000-0001-7616-1554]{E.~Aakvaag}$^\textrm{\scriptsize 17}$,
\AtlasOrcid[0000-0002-5888-2734]{B.~Abbott}$^\textrm{\scriptsize 121}$,
\AtlasOrcid[0000-0002-0287-5869]{S.~Abdelhameed}$^\textrm{\scriptsize 83b}$,
\AtlasOrcid[0000-0002-1002-1652]{K.~Abeling}$^\textrm{\scriptsize 54}$,
\AtlasOrcid[0000-0001-5763-2760]{N.J.~Abicht}$^\textrm{\scriptsize 48}$,
\AtlasOrcid[0000-0002-8496-9294]{S.H.~Abidi}$^\textrm{\scriptsize 30}$,
\AtlasOrcid[0009-0003-6578-220X]{M.~Aboelela}$^\textrm{\scriptsize 44}$,
\AtlasOrcid[0000-0002-9987-2292]{A.~Aboulhorma}$^\textrm{\scriptsize 36e}$,
\AtlasOrcid[0000-0001-5329-6640]{H.~Abramowicz}$^\textrm{\scriptsize 154}$
% ...
\bigskip
\\

% ...
$^{17}$Department for Physics and Technology, University of Bergen, Bergen; Norway.\\
% ...
$^{30}$Physics Department, Brookhaven National Laboratory, Upton NY; United States of America.\\
% ...
$^{36}$$^{(a)}$Facult\'e des Sciences Ain Chock, Universit\'e Hassan II de Casablanca;$^{(b)}$Facult\'{e} des Sciences, Universit\'{e} Ibn-Tofail, K\'{e}nitra;$^{(c)}$Facult\'e des Sciences Semlalia, Universit\'e Cadi Ayyad, LPHEA-Marrakech;$^{(d)}$LPMR, Facult\'e des Sciences, Universit\'e Mohamed Premier, Oujda;$^{(e)}$Facult\'e des sciences, Universit\'e Mohammed V, Rabat;$^{(f)}$Institute of Applied Physics, Mohammed VI Polytechnic University, Ben Guerir; Morocco.\\
% ...
$^{44}$Physics Department, Southern Methodist University, Dallas TX; United States of America.\\
% ...
$^{48}$Fakult\"{a}t Physik, Technische Universit{\"a}t Dortmund, Dortmund; Germany.\\
% ...
$^{54}$II. Physikalisches Institut, Georg-August-Universit\"{a}t G\"ottingen, G\"ottingen; Germany.\\
% ...
$^{83}$$^{(a)}$Khalifa University of Science and Technology, Abu Dhabi;$^{(b)}$New York University Abu Dhabi, Abu Dhabi;$^{(c)}$United Arab Emirates University, Al Ain;$^{(d)}$University of Sharjah, Sharjah; United Arab Emirates.\\
% ...
$^{102}$CPPM, Aix-Marseille Universit\'e, CNRS/IN2P3, Marseille; France.\\
% ...
$^{121}$Homer L. Dodge Department of Physics and Astronomy, University of Oklahoma, Norman OK; United States of America.\\
% ...
$^{154}$Raymond and Beverly Sackler School of Physics and Astronomy, Tel Aviv University, Tel Aviv; Israel.\\
% ...
$^{a}$ Also at Affiliated with an institute formerly covered by a cooperation agreement with CERN.\\
$^{b}$ Also at An-Najah National University, Nablus; Palestine.\\
$^{c}$ Also at Borough of Manhattan Community College, City University of New York, New York NY; United States of America.\\
$^{d}$ Also at Center for Interdisciplinary Research and Innovation (CIRI-AUTH), Thessaloniki; Greece.\\
$^{e}$ Also at Centre of Physics of the Universities of Minho and Porto (CF-UM-UP); Portugal.\\
$^{f}$ Also at CERN, Geneva; Switzerland.\\
% ...

\end{flushleft}


%

\clearpage

\end{document}